\documentclass[twocolumn,showpacs,preprintnumbers,amsmath,prl,amssymb,superscriptaddress]{revtex4}

\usepackage{graphicx}% Include figure files
\usepackage{dcolumn}% Align table columns on decimal point
\usepackage{bm}% bold math

\def\cpkkd{\rm{kg^{-1}keV^{-1}day^{-1}}}
\def\mwimp{\rm{m_{\chi}}}
\def\csnospin{\rm{\sigma_{\chi N}^{SI}}}
\def\csspin{\rm{\sigma_{\chi n}^{SD} }}
\def\munu{\mu_{\nu}}
\def\sasix{\rm{SA_{6}}}
\def\sa12{\rm{SA_{12}}}
\def\trigeff{\epsilon _{\rm Trig}}
\def\daqeff{\epsilon _{\rm DAQ}}
\def\anlyeff{\epsilon _{\rm Anly}}
\def\psdeff{\epsilon _{\rm PSD}}

\begin{document}

\preprint{AS-TEXONO/07-06}

\title{
New limits on spin-independent 
and spin-dependent couplings of 
low-mass WIMP dark matter with a 
germanium detector at a threshold 
of 220~eV
}

%%Begin Author List
%Authorlist v.2.8 May 6, 2002
%

\newcommand{\as}{Institute of Physics, Academia Sinica, Taipei 115, Taiwan.}
\newcommand{\thu}{Department of Engineering Physics, Tsinghua University,
Beijing 100084, China.}
\newcommand{\metu}{Department of Physics,
Middle East Technical University, Ankara 06531, Turkey.}
\newcommand{\ihep}{Institute of High Energy Physics,
Chinese Academy of Science, Beijing 100039, China.}
\newcommand{\ciae}{Department of Nuclear Physics,
Institute of Atomic Energy, Beijing 102413, China.}
\newcommand{\bhu}{Department of Physics, Banaras Hindu University,
Varanasi 221005, India.}
\newcommand{\corr}{htwong@phys.sinica.edu.tw;
Tel:+886-2-2789-6789; FAX:+886-2-2788-9828.}

\author{ S.T.~Lin }  \affiliation{ \as }
\author{ H.B.~Li }  \affiliation{ \as }
\author{ X.~Li } \affiliation{ \thu }
\author{ S.K.~Lin }  \affiliation{ \as }
\author{ H.T.~Wong } \altaffiliation[Corresponding Author: ]{ \corr } \affiliation{ \as }
\author{ M.~Deniz } \affiliation{ \as } \affiliation{ \metu }
\author{ B.B.~Fang } \affiliation{ \thu }
\author{ D.~He } \affiliation{ \thu }
\author{ J.~Li }  \affiliation{ \thu } \affiliation{ \ihep }
\author{ C.W.~Lin }  \affiliation{ \as }
\author{ F.K.~Lin }  \affiliation{ \as }
\author{ X.C.~Ruan } \affiliation{ \ciae }
\author{ V.~Singh }  \affiliation{ \as } \affiliation{ \bhu }
\author{ A.K.~Soma }  \affiliation{ \as } \affiliation{ \bhu }
\author{ J.J.~Wang }  \affiliation{ \as }
\author{ Y.R.~Wang } \affiliation{ \as }
\author{ S.C.~Wu } \affiliation{ \as }
\author{ Q.~Yue } \affiliation{ \thu }
\author{ Z.Y.~Zhou } \affiliation{ \ciae }

\collaboration{TEXONO Collaboration}

%%\noaffiliation

%%End Author List

\date{\today}% It is always \today, today,

\begin{abstract}

An energy threshold of (220$\pm$10)~eV 
was achieved at an efficiency of 50\%
with a four-channel ultra-low-energy germanium detector
each with an active mass of 5~g.
This provides a unique probe to 
WIMP dark matter with mass below 10~GeV.
With a data acquisition live time of 0.338~kg-day
at the Kuo-Sheng Laboratory,
constraints on WIMPs in the galactic halo 
were derived.
The limits improve over previous results 
on both spin-independent WIMP-nucleon
and spin-dependent WIMP-neutron cross-sections 
for WIMP mass between 3$-$6~GeV.
Sensitivities for full-scale experiments are projected.
This detector technique makes
the unexplored sub-keV energy window 
accessible for new neutrino and dark matter experiments.

\end{abstract}

% PACS, the Physics and Astronomy % Classification Scheme.
%Use showkeys class option if keyword %display desired

\maketitle

There is compelling evidence from cosmological and
astrophysical observations 
that about one quarter of
the energy density of the universe 
can be  attributed to
Cold Dark Matter(CDM), whose nature 
and properties are still unknown\cite{pdgcdm}.
Weakly Interacting Massive Particles 
(WIMP, denoted by $\chi$) are
the leading candidates for CDM.
There are intense experimental efforts\cite{wimpexpt}
to look for WIMPs through direct detection 
of nuclear recoils  in
$\chi$N$\rightarrow$$\chi$N 
elastic scattering or in the studies of the possible
products through $\chi \bar{\chi}$ 
annihilations.

Supersymmetric(SUSY) particles\cite{pdgsusy} 
are the leading WIMP candidates.
The popular SUSY models prefer WIMP mass($\mwimp$) in the range
of $\sim$100~GeV,
though light neutralinos 
remain a possibility\cite{lightsusy}.
Most experimental programs
optimize their design in the high-mass region
and exhibit diminishing sensitivities for $\rm{\mwimp < 10 ~GeV}$,
where an allowed region 
due to the annual modulation data
of the DAMA experiment\cite{damaallowed,damasdep}
$-$ further reinforced by the first 
DAMA/LIBRA results\cite{damalibra} $-$ 
remains unprobed. 
Simple extensions of the Standard Model with a 
singlet scalar favors light WIMPs\cite{smscalar}.
Detectors with sub-keV threshold are needed 
for probing this low-mass region
and studying WIMPs bound in the solar system\cite{solarwimp},
and non-pointlike SUSY candidates like Q-balls\cite{qball}.
This presents a formidable
challenge to both detector technology and background control.
Only the CRESST-I experiment has set limits\cite{cresst1} 
with sapphire($\rm{Al_2 O_3}$)-based cryogenic detector
at a threshold of 600~eV.

The Kuo-Sheng(KS) Laboratory\cite{texonoprogram}
is located at 28~m from a 2.9~GW 
reactor core with an overburden of about 30~meter-water-equivalence.
Limits on neutrino magnetic moments($\munu$)\cite{munureview} 
with a 1.06-kg germanium detector(HPGe) at a threshold of
5~keV were reported\cite{texonomagmom}. 
These data also allowed
the studies of reactor electron neutrinos\cite{rnue}
and reactor axions\cite{raxion}.
A background level  of
${\rm \sim 1 ~ event ~ \cpkkd  }$(cpkkd) at 20~keV,
comparable with those of underground CDM experiments,
was achieved.
The current goal is to 
develop %%advanced 
detectors with
kg-scale target mass, 100~eV-range threshold
and low-background specifications
for the studies of WIMPs, $\munu$ and
neutrino-nucleus coherent scatterings\cite{texonocohsca}.

Ultra-low-energy germanium detectors(ULEGe)
is a matured technique for 
sub-keV soft X-rays measurements.  %% in material and biophysics. 
They typically have modular mass of
5$-$10~g while detector arrays of up to 30 elements
have been constructed.
Compared with $\rm{Al_2 O_3}$,
Ge provides enhancement in 
$\chi$N spin-independent couplings($\csnospin$)
due to the $\rm{ A^2 }$ dependence\cite{pdgcdm,cdmmaths}, 
where A is the mass number of the target isotopes.
%%In addition, 
The isotope $^{73}$Ge (natural isotopic
abundance of 7.73\%)  
comprises an unpaired neutron such that
it can provide additional probe to 
the spin-dependent couplings of WIMPs with 
the neutrons($\csspin$).
The nuclear recoils from $\chi$N interactions
in ULEGe
only give rise to $\sim$20\% of the observable ionizations
compared with electron recoils at the same energy.
The suppression ratio is called the  
quenching factor(QF)\cite{qftheory}.
For clarity, all ULEGe measurements discussed
hereafter in this article are 
electron-equivalent-energy,
unless otherwise stated. 

\begin{figure}[hbt]
\includegraphics[width=7.5cm]{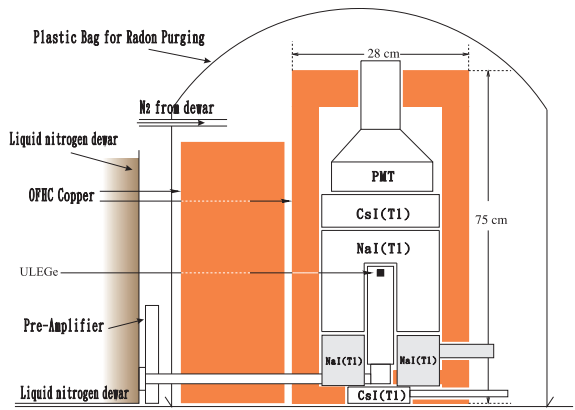}
\caption{
Schematic layout of the ULEGe
with its anti-Compton detectors,
inner shieldings and
radon purge system.
A 50-ton shielding 
structure\cite{texonomagmom} 
encloses the detectors.
}
\label{ksulege}
\end{figure}

The ULEGe array consists of 
4-element 
each having an active mass of 5~g\cite{ulege}. 
Standard ultra-low-background
specifications were adopted in its
construction and
choice of materials.
It has identical external dimensions
as the 1-kg HPGe of Ref.~\cite{texonomagmom}.
Apart from swapping between the two detectors,
data taking was performed
with all other hardware components,
shieldings configuration,
electronics and data acquisition (DAQ) systems\cite{eledaq}
kept identical.
The schematic diagram of the experimental setup
inside the shieldings is depicted in Figure~\ref{ksulege}.

The ULEGe signals were provided by built-in 
pulsed optical feedback pre-amplifiers,
and were distributed to two spectroscopy
amplifiers at 6~$\mu$s($\sasix$) 
and 12~$\mu$s($\sa12$) shaping times
and with different amplification factors.
Discriminator output of $\sasix$
defined the trigger conditions for DAQ.
The threshold was set to about 4.3$\pm$0.2 times the
RMS fluctuations of the $\sasix$ signals 
above the pedestal.
The DAQ rates for the ULEGe were about 5~Hz,
due mostly to electronic noise and agreed well
with expectations\cite{collar0806,statham}.
The $\sasix$, $\sa12$, Anti-Compton Veto (ACV) 
and Cosmic-Ray Veto (CRV) signals 
were read out by 20~MHz 
Flash Analog-to-Digital Convertors.
Random trigger(RT) events 
generated at 0.1~Hz and
uncorrelated with the rest of the system,
as well as various system control parameters,
were also recorded.

\begin{figure}
\includegraphics[width=7.5cm]{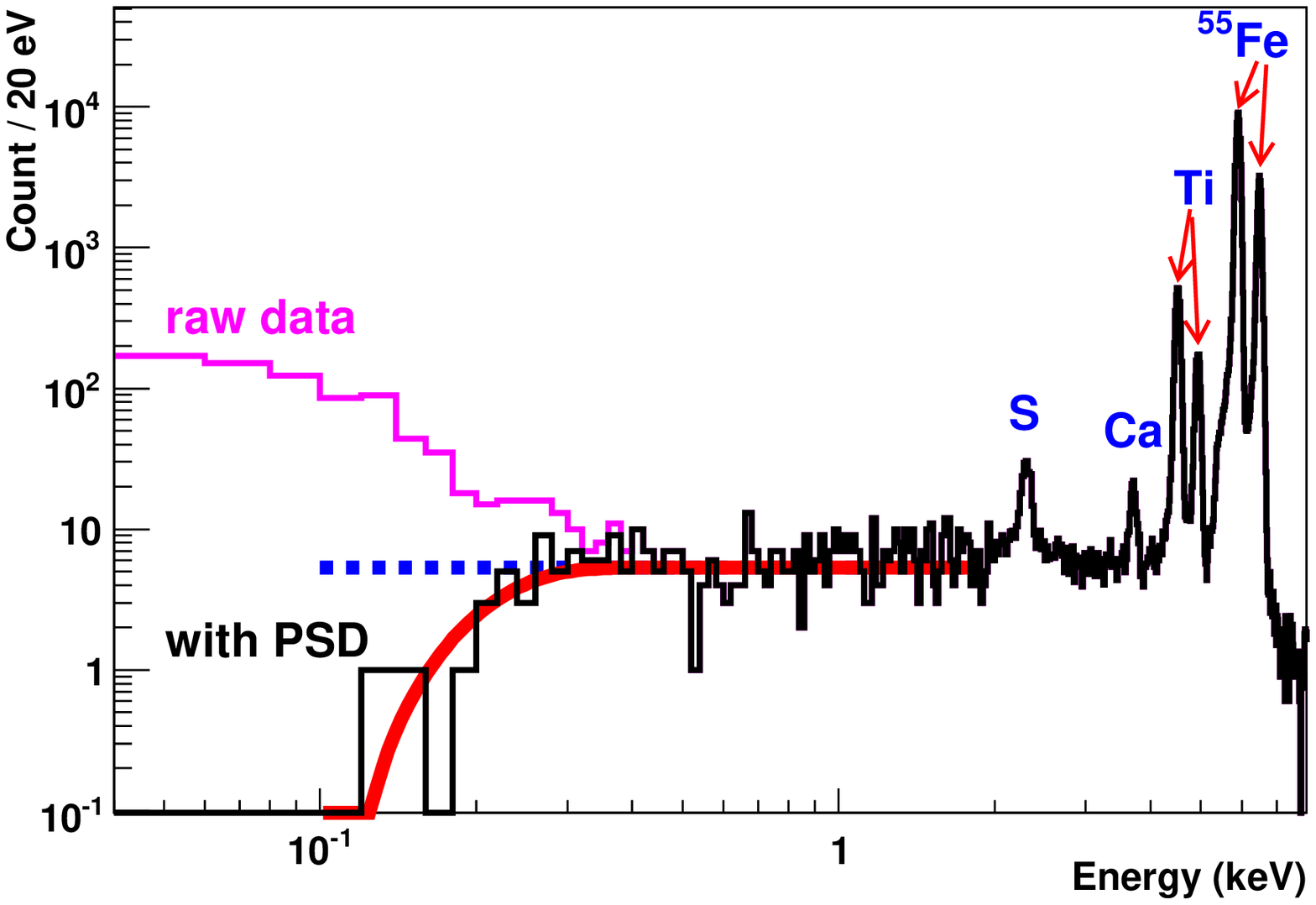}\\[0.5cm]
\includegraphics[width=7.5cm]{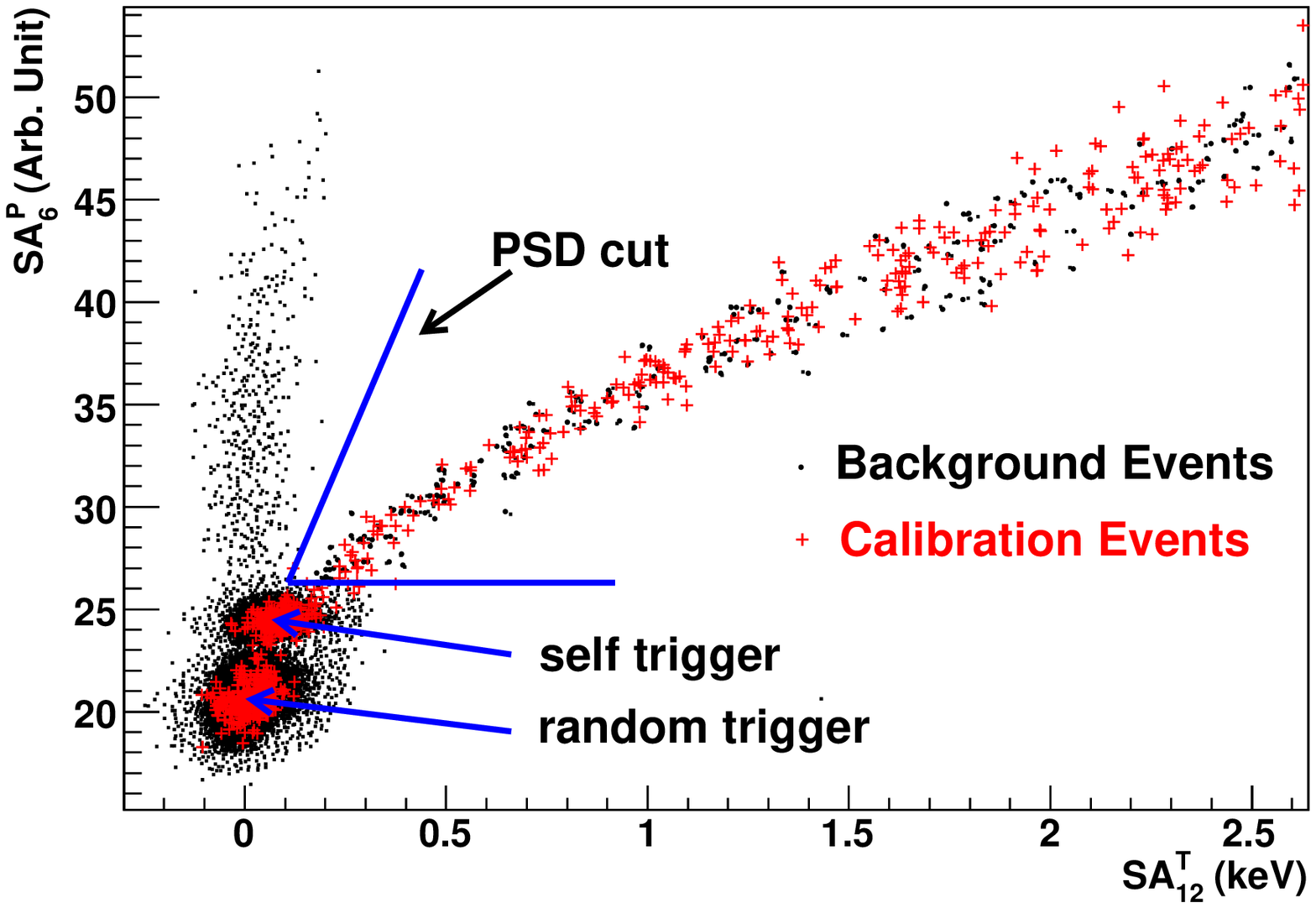}
\caption{
(a) Top:
Measured energy spectrum of the ULEGe
with $^{55}$Fe
source together with X-ray peaks 
from various materials.
The black histogram represents events 
selected by PSD cuts.
Deviations from the expected flat spectra
contribute to PSD efficiencies.
(b) Bottom:
Scattered plots 
of the $\rm{SA^{P}_{6}}$ versus
$\rm{SA^T_{12}}$ signals,
for both calibration and physics events.
The PSD selection is shown.
}
\label{fe55}
\end{figure}

Energy measurements were given by 
$\rm{SA_{12}^T}$ defined in the next paragraph.
Figure~\ref{fe55}a
shows an energy calibration spectrum due to
external $^{55}$Fe sources(5.90 and  6.49~keV)
together with X-rays from
Ti(4.51 and 4.93~keV), Ca(3.69~keV),
and S(2.31~keV).
Photons with energy lower than 2~keV were completely
absorbed by the detector window.
The RT-events provided the calibration
point at zero-energy.
The RMS resolutions for the 
RT-events and $^{55}$Fe peaks
were about 55~eV and 78~eV, respectively.
The calibration procedures were performed before
and after the DAQ periods. 
Linearity was checked up to
60~keV with various $\gamma$-sources, and
between zero and 2~keV with a 
precision pulse generator.
The energy scale was accurate to 
${\rm < 20~eV}$, while deviations from
linearity were  ${\rm < 1\%}$.
The electronic gain drifts, also 
monitored {\it in situ} by the
pulse generator, were ${\rm < 5\%}$.
A detector hardware ``noise-edge'' of
about 300~eV was achieved.

%%The analysis procedures followed
%%closely to those of $\munu$ studies\cite{texonomagmom}.
Pulse shape discrimination (PSD) software 
was devised to differentiate physics events 
from those due to electronic noise,
exploiting the correlations in
both the energy and timing
information of the $\sasix$ and $\sa12$ signals.
Displayed in Figure~\ref{fe55}b
is a scattered plot 
of the $\rm{SA^{P}_{6}}$ and 
$\rm{SA^T_{12}}$ signals
with the PSD cut superimposed, where
the superscripts P/T denote partial/total
integration of the pulses 
within 
\mbox{(15,25)}~$\mu$s
and 
\mbox{(-20,52)}~$\mu$s, 
respectively,
relative to the trigger instant(t=0).
The noise events were suppressed. 
Calibration events and
those from physics background 
were overlaid, indicating 
uniform response.
Events selected by PSD but with CRV or ACV tags 
were subsequently rejected.
The surviving events were ULEGe signals uncorrelated 
with other detector systems
and could be WIMP candidates.

\begin{figure}
\includegraphics[width=7.5cm]{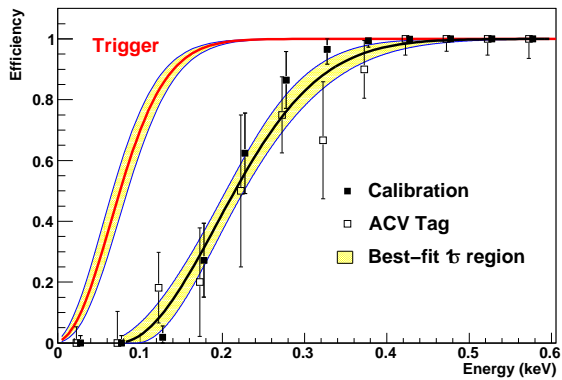}
\caption{
Selection efficiencies of the PSD cut, as
derived from the $^{55}$Fe-calibration
and {\it in situ} data with ACV tags.
Also shown are 
the best-fit 1$\sigma$ region and
the trigger efficiency 
for physics events recorded by the DAQ
system.
}
\label{eff}
\end{figure}

\begin{table*}
\caption{
\label{summarytable}
Summary of analysis results
and their statistical
and systematic errors in
the two energy intervals
which define the sensitivities at
low and high WIMP masses.
}
\begin{ruledtabular}
\begin{tabular}{lcc}
Energy Bin & 198$-$241~eV  & 1.39$-$1.87~keV \\ \hline
Raw Background Counts & 105212  & 75 \\
\underline{Selection Cuts and Systematic Effects}~: & & \\
~~ Trigger Efficiency (\%) & \multicolumn{2}{c}{ 100 } \\
~~ DAQ Dead Time (\%) & \multicolumn{2}{c}{ $\rm{ 11.0 \pm 0.1 }$} \\
~~ PSD $-$ Cumulative Background Survival Fraction (\%) 
& 0.008  &  97 \\
\hspace*{1.4cm} Signal Efficiency (\%) & 
\multicolumn{2}{c}{$\rm{ 66 \pm 6 }$ \hspace*{1cm}  100 ~~~} \\ 
~~ ACV $-$ Cumulative Background Survival Fraction  (\%)
& 0.0  &  2.7 \\
\hspace*{1.4cm} Signal Efficiency (\%) & 
\multicolumn{2}{c}{$ \rm{ 98.3 \pm 0.1 }$ } \\
~~ CRV $-$ Cumulative Background Survival Fraction (\%)
& 0.0  &  0.0 \\
\hspace*{1.4cm} Signal Efficiency (\%) & 
\multicolumn{2}{c}{$ \rm{ 91.4 \pm 0.1 }$ } \\ \hline
After-Cut Background Counts & 0 & 0 \\
After-Cut Normalized Background Rates ($\cpkkd$)
& $\rm{ 0 \pm ^{272}_{~~0} (stat) \pm ^{30}_{27} (sys)}$   &
$\rm{ 0 \pm ^{13}_{~0} (stat) \pm 0 (sys) }$  \\ 
Quenching Factor & 
$\rm{ 0.200 \pm 0.006}$  & $\rm{ 0.244 \pm 0.007}$  \\ \hline
Sampling in $\mwimp$ (GeV)  & 5 & 50  \\
\underline{$\csnospin ~ {\rm ( 10^{-39} ~ cm^2 )} $} :  & & \\
~~ Mean \& Errors due to Background \& QF Uncertainties &  
$\rm{ 0  \pm ^{0.64}_{~~~0} (Bkg) \pm  {0.01} (QF)}$   &
$\rm{ 0 \pm ^{0.153}_{~~~~0} (Bkg) \pm {0.003} (QF)}$   \\
~~ Limit at 90\% Confidence Level   
&  $\rm{ < 0.81 }$  & $\rm{ < 0.20 }$  \\
\underline{$\csspin ~ {\rm ( 10^{-34} ~ cm^2 )} $} :  & & \\
~~ Mean \& Errors due to Background \& QF Uncertainties &  
$\rm{ 0  \pm ^{1.90}_{~~~0} (Bkg) \pm  {0.03} (QF)}$   &
$\rm{ 0 \pm ^{0.47}_{~~~0} (Bkg) \pm {0.01} (QF)}$   \\
~~ Limit at 90\% Confidence Level   
&  $\rm{ < 2.40 }$  & $\rm{ < 0.59 }$  \\
\end{tabular}
\end{ruledtabular}
\end{table*}

Data were taken with the ULEGe at KS
with different hardware and software
configurations.
They provided important input 
on the background understanding 
and performance optimizations 
for future full-scale experiments.
The data set with
the best background and threshold
has a DAQ live time of 0.338~kg-day. 
The analysis results and the
systematic effects at the two energy intervals
which defined the sensitivities for
$\mwimp$ below and above $\sim$10~GeV
are summarized in Table~\ref{summarytable}.

\begin{figure}
\includegraphics[width=7.5cm]{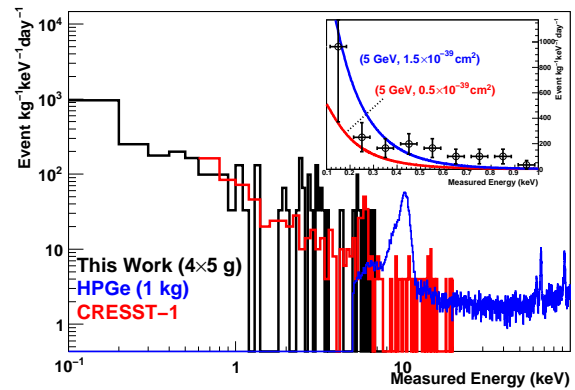}
\caption{
The measured spectrum of ULEGe with
0.338~kg-day of data,
after CRV, ACV and PSD selections.
Background spectra of the CRESST-I experiment\cite{cresst1}
and the HPGe\cite{texonomagmom}
are overlaid for comparison.
The expected nuclear recoil spectra
for two cases
of $\rm{ ( \mwimp , \csnospin ) }$
are superimposed onto the spectrum
shown in linear scales in the inset.
}
\label{bkgspect}
\end{figure}

The ULEGe data were taken in conjunction with a
CsI(Tl) scintillator array\cite{kscsi} for
the studies of neutrino-electron scattering.
The combined DAQ rate was about 30~Hz.
The DAQ dead time
and the CRV+ACV selection efficiencies
listed in Table~\ref{summarytable}
were accurately measured using
RT-events\cite{texonomagmom}.
The maximum amplitude distributions of 
the RT pedestals and physics events 
above the noise-edge
of 300~eV were measured.
The corresponding distributions
for events $<$300~eV  were evaluated by
interpolations to avoid biased sampling.
The trigger efficiencies
depicted in Figure~\ref{eff}
correspond to
the fractions of the distributions
above the discriminator threshold level. 

Events in coincidence with ACV-tags 
are mostly physics-induced.
The fraction of these events surviving 
the PSD cuts was taken to be 
the PSD efficiency.
This assignment is conservative since
the actual efficiency corresponds to
the survival fraction of
samples {\it after} electronic noise events 
in accidental coincidence 
were subtracted, 
and therefore should be higher.
Alternatively,
under the assumption that 
the $^{55}$Fe calibration of Figure~\ref{fe55}a
would give rise to physics events with
a flat spectrum down to the lowest
energy relevant to this analysis ($<$100~eV),
the deviations of the PSD-selected events
from a flat distribution
provided the second measurement.
Consistent results were obtained with both approaches,
as depicted in Figure~\ref{eff}.
The larger uncertainties of the first method are 
due to the limited statistics from only
the {\it in situ} ACV samples.
The efficiencies and their uncertainties
adopted for analysis were
derived from a best-fit on the combined data set.
A threshold of (220$\pm$10)~eV  was achieved with
a PSD efficiency of 50\%.

\begin{figure}
\includegraphics[width=8cm]{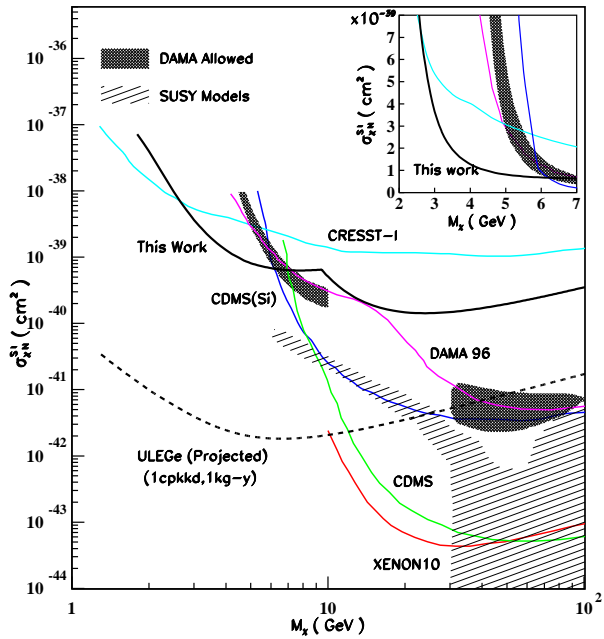}
\caption{
Exclusion plot of the
spin-independent $\chi$N
cross-section versus
WIMP-mass, displaying the
KS-ULEGe limits and
those defining the
current boundaries\cite{cresst1,cdmbounds}.
The DAMA-allowed regions\cite{damaallowed}
are superimposed.
The striped region is that favored
by SUSY models\cite{lightsusy}.
Projected sensitivities of full-scale
experiments are indicated as dotted lines.
}
\label{explotsindep}
\end{figure}

The ULEGe spectrum normalized in cpkkd unit
after the CRV, ACV and PSD selections
is displayed in Figure~\ref{bkgspect},
showing comparable background
as CRESST-I\cite{cresst1}. %%with 1.51~kg-day of data,
Listed in Table~\ref{summarytable} are
the normalized background rates,
indicating that statistical uncertainties
dominate over the systematic effects.
%%The standard and conservative approach
%%that 
%%No background profile was subtracted such
%%that the WIMP signals cannot be larger than the
%%observed event rates.
%% was adopted.
The formalisms followed those of Ref.~\cite{cdmmaths}
using standard nuclear form factors,
a galactic rotational velocity of
${\rm 230 ~ km \, s^{-1}}$ and 
a local WIMP density of $\rm{0.3 ~ GeV \, cm^{-3}}$ 
with Maxwellian velocity distribution.
No subtraction of background profiles was made 
such that the WIMP signals cannot be larger than the
observed event rates.
The unbinned optimal interval method 
as formulated in Ref.~\cite{yellin}
and widely used by current CDM experiments
was adopted to derive the upper
limits for the possible $\chi$N event rates.
By comparing the observed background in 
different energy intervals 
with the expected number of events due to
$\chi$N recoils for each $\mwimp$,
the optimal intervals 
producing the most stringent limits to 
$\csnospin$ and $\csspin$ were selected.
Corrections due to QF, detector resolution
and various efficiency factors 
were incorporated.
The energy dependence of QF in Ge was
evaluated with the TRIM software package\cite{trim}.
The uncertainties were taken to be their
difference with the statistical best-fit values 
of the available data\cite{geqf} 
from 254~eV to 200~keV nuclear recoil energy.

\begin{figure*}[hbt]
\includegraphics[width=8cm]{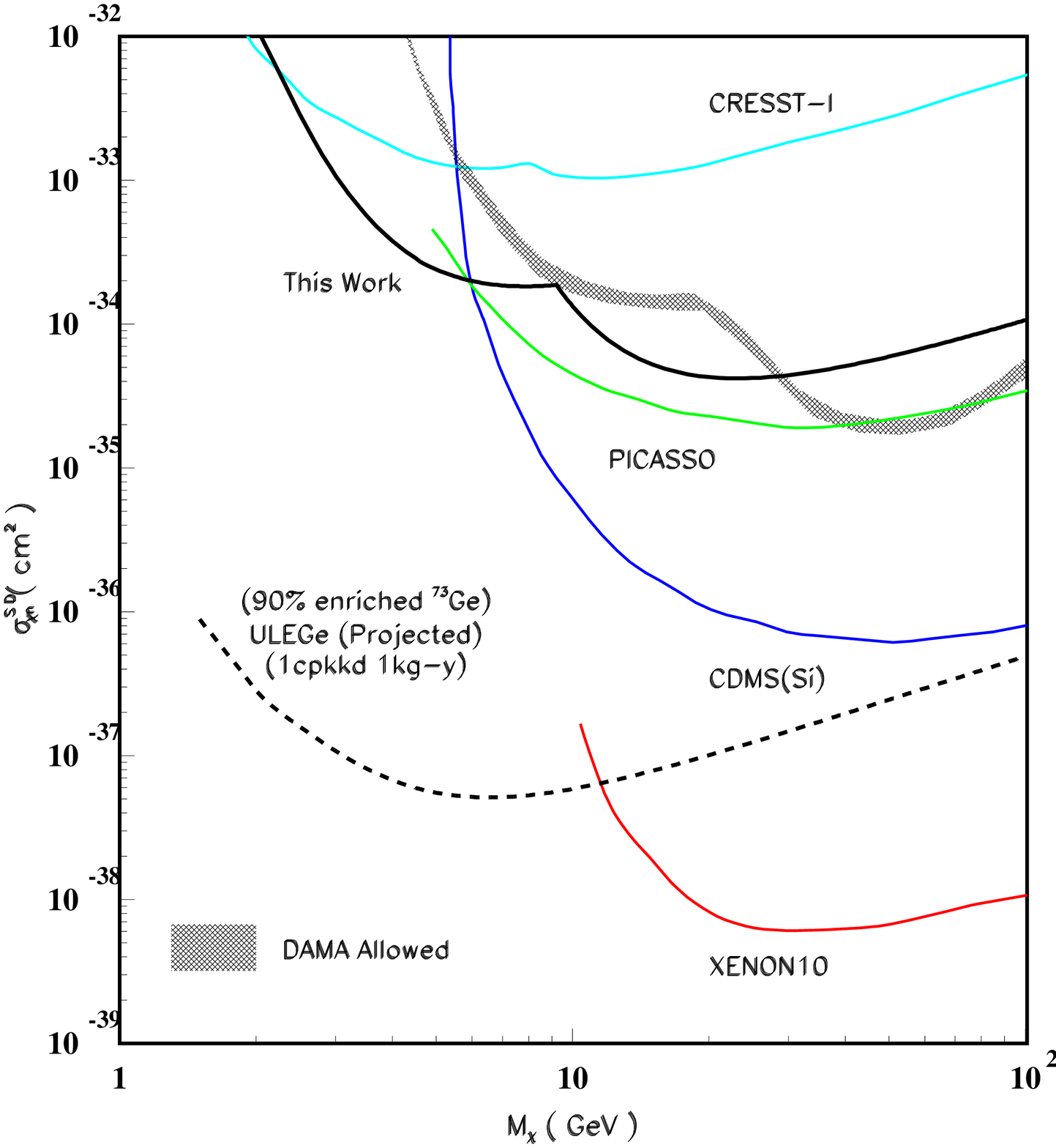}
\hspace*{0.5cm}
\includegraphics[width=8cm]{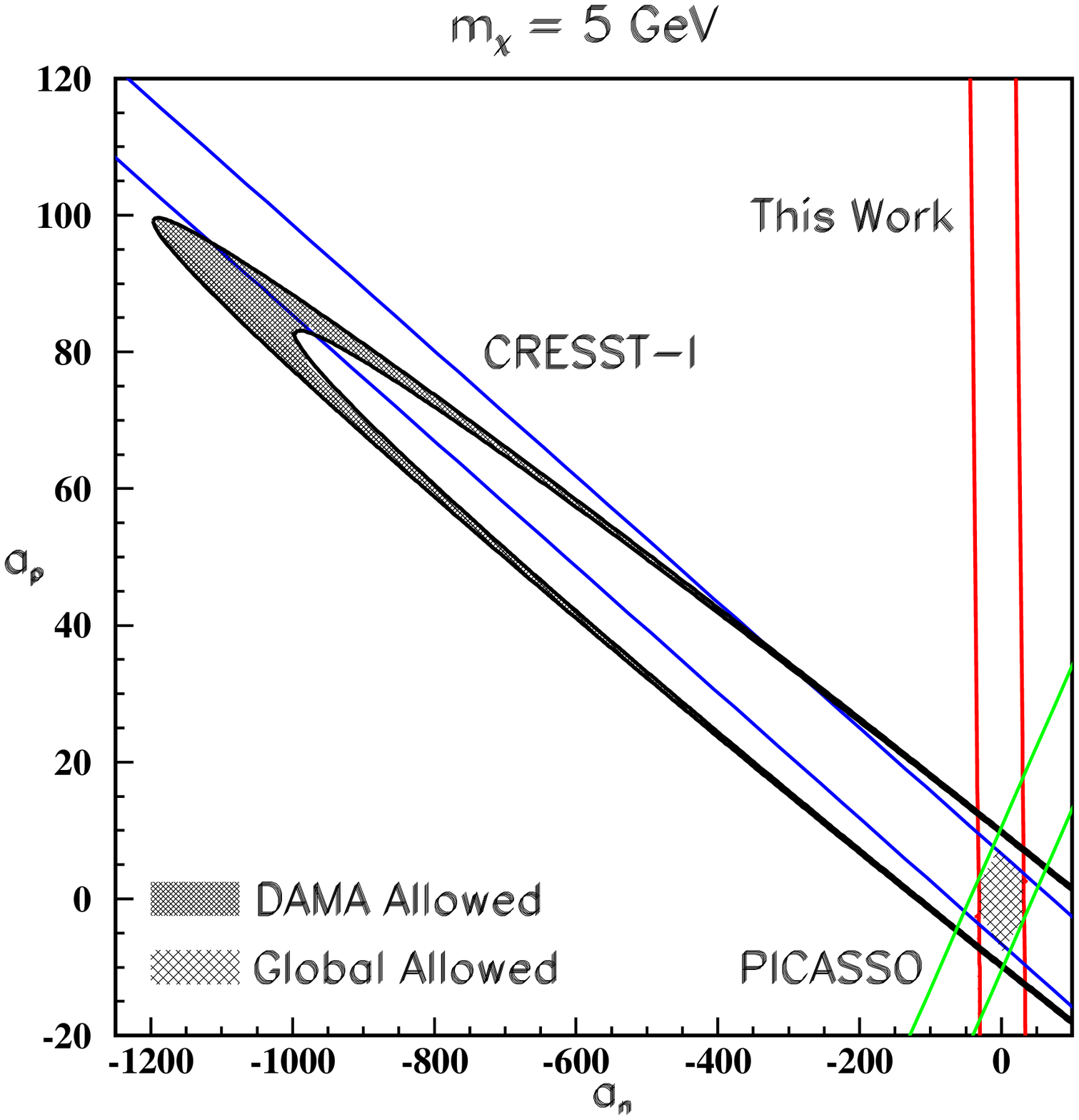}
\caption{
(a) Left:
Exclusion plot of the
spin-dependent $\chi$-neutron 
cross-section versus
WIMP-mass. Same conventions
as those in Figure~\ref{explotsindep} are used.
(b) Right:
Constraints at ${\rm \mwimp = 5 ~ GeV}$
on the effective
axial four-fermion
$\chi$-proton ($a_p$) and
$\chi$-neutron ($a_n$)
spin-dependent couplings, in units
of $\rm{ 2 \sqrt{2} G_F}$
following the formulation of Ref.~\cite{sdeptovey}.
The parameter space within the bands
of the corresponding experiments are allowed.
The shaded area at the origin
is the combined allowed region.
}
\label{explotsdep}
\end{figure*}

Exclusion plots
on both $( \mwimp ,  \csnospin )$ 
and $( \mwimp ,  \csspin )$ planes
at 90\% confidence level
for galactically-bound WIMPs
were then derived, as
depicted in Figures~\ref{explotsindep}
and \ref{explotsdep}a, respectively.
The DAMA-allowed regions\cite{damaallowed,damasdep}
and the current exclusion boundaries\cite{cresst1,cdmbounds}
are displayed.
The model-independent approach
of Refs.~\cite{sdepbottino,sdeptovey}
were adopted to extract limits on the
spin-dependent cross-sections.
Consistent results were obtained when
different $^{73}$Ge nuclear physics 
matrix elements\cite{ge73np} were adopted
as input.
The constraints on the effective
axial four-fermion
$\chi$-proton and $\chi$-neutron
spin-dependent couplings\cite{sdeptovey}
at $\mwimp$=5~GeV
are displayed in Figure~\ref{explotsdep}b.
The parameter space probed by the $^{73}$Ge in ULEGe
is complementary
to that of the CRESST-I experiment\cite{cresst1}
where the $^{27}$Al target is made up
of an unpaired proton instead.
New limits were set by
the KS-ULEGe data in both
$\csnospin$ and $\csspin$
for $\mwimp$$\sim$3$-$6 ~GeV.
The remaining DAMA low-$\mwimp$ allowed regions
in both interactions were
probed and excluded.
The observable nuclear recoils
at $\mwimp$=5~GeV and
$\rm\csnospin$=${\rm 0.5 \times 10^{-39} ~ cm^2}$(allowed)
and ${\rm 1.5 \times 10^{-39} ~ cm^2}$(excluded)
are superimposed with the measured spectrum
in the inset of Figure~\ref{bkgspect} for illustrations.
It is expected that
recent data from the COUPP\cite{coupp08} experiment 
can place further constraints in the 
spin-dependent plots of 
Figures~\ref{explotsdep}a\&b.

This work extends the bounds on WIMPs 
by making measurements 
in a new observable window of 100~eV$-$1~keV
in a low-background environment.
Understanding and suppression
of background at this sub-keV region is
crucial for further improvement.
Measurements are conducted with
the ULEGe at an underground laboratory.
There are recent advances in 
``Point-Contact'' Ge detector\cite{chicago} 
which offer potentials of scaling-up the detector
mass to the kg-range.
Preliminary results in dark matter searches  
were recently reported\cite{cogent08}.
The mass-normalized external
background will be reduced in 
massive detectors due to self-attenuation\cite{texonocohsca}.
Further reduction in threshold may be possible
with improved junction field-effect-transistors and
by correlating signals from both electrodes.
The potential reach of full-scale experiments with
1~kg-year of data and a benchmark background level of 
1~cpkkd is illustrated in Figures~\ref{explotsindep}
and \ref{explotsdep}a.
Such experimental programs 
are complementary
to the many current efforts on 
CDM direct searches. 

We are grateful 
to the KIMS Collaboration
and the authors of Ref.~\cite{collar0806}
for inspiring comments.
This work is supported by
the Academia Sinica Pilot Project 2004-06,
Theme Project 2008-10,
contracts 95-2119-M-001-028 and 96-2119-M-001-005
from the National Science Council, Taiwan
and 10620140100 from the 
National Natural Science Foundation, China.

% ****** Start of file wimp-reply-v2.tex ****** 
% 

%%%%%%%%%%%%%%%%%%%%%%%%%%%%%%%%%%%%%%%%%%%%%%%%%%%%%

\newpage

\setcounter{figure}{0}

\begin{center}

\large
{\bf 
Appendix I :
\vspace*{0.2cm}

Trigger Efficiency at Threshold $-$\\
Reply to arXiv:0806.1341v2 
}
%%}

\normalsize

\vspace*{0.2cm}

H.B.~Li, S.T.~Lin, S.K.~Lin, A.K.~Soma, H.T.~Wong$^*$\\
(TEXONO Collaboration)\\
$^*$ Corresponding Author: htwong@phys.sinica.edu.tw

\end{center}

\vspace*{0.2cm}

The authors
in a recent paper~\cite{collar0806v2a1}
raised questions
on our estimates of trigger efficiency in 
the evaluation of the constraints on WIMP Dark Matter in the 
low-mass ($< 10 ~ {\rm GeV}$) domain with a
Ultra-Low Energy Germanium (ULEGe) detector at 
a threshold of 220~eV~\cite{texono0712a1}.
The discrepancy originates from  
some misleading terminology in Ref.~\cite{texono0712a1}.
We address the issue in details in this reply.
We  show how the trigger efficiencies are
derived and demonstrate that 
the results  of Ref.~\cite{texono0712a1}
are correct.

\section{I. Concept of Experiment}

Details of the experimental setup and
data analysis, as well as the definitions
of notations, can be referred to Ref.~\cite{texono0712a1}.
There are two categories of events
relevant to the present discussion:
physics signals (PHY) and electronic noise (ELE).
The PHY events are due to actual energy depositions
at the ULEGe by gammas, neutrons, neutrinos, WIMPs 
and other radiations, while ELE events are induced by the various
stages in the readout electronics.
The majority of the events above the noise edge
of 300~eV are PHY-events. 
A particularly good method to extract a clean sample 
of PHY-events below the noise edge for further studies
is to require that the ULEGe signals are in coincidence
with the Anti-Compton Veto (ACV) tag.
However, PHY-events due to neutrinos or dark matter
interactions would have to be in anti-coincidence with ACV.
A major goal of the experiment is to lower the
threshold by suppressing ELE while keeping PHY-events
in some substantially large and known fraction.

In order for PHY-events to be included  in the final
spectra where physics is extracted,
they have to survive three procedures, the 
efficiency of each of which must be known.
These efficiency factors were summarized in Table~1
of Ref.~\cite{texono0712a1}:
\begin{enumerate}
\item 
Trigger efficiency ($\trigeff$) $-$
that
PHY would produce a trigger signal to the
data acquisition (DAQ) system. This efficiency
depends on the energy of PHY.
\item 
DAQ efficiency ($\daqeff$) $-$
that
the trigger signal would actually activate
the DAQ system resulting in a complete event recorded in the
computer. This efficiency is independent of the
energy of PHY.
\item 
Analysis efficiency ($\anlyeff$) $-$
that
offline software procedures would retain PHY and suppress ELE.
Some PHY would be rejected in the process leading to
an efficiency factor which is energy-dependent.
\end{enumerate}

Among these three efficiency factors, 
the derivations of $\daqeff$ and
$\anlyeff$ were discussed in Ref.~\cite{texono0712a1}.
The $\daqeff$ was evaluated accurately
by ``Random Trigger'' (RT) events to be 89\%
-- fraction of the RT events actually recorded in the computer.
(cited in Table~1 as ``DAQ Dead Time'' of 11.0\%.)
Same procedures were used in our earlier work on neutrino
magnetic moments~\cite{texonomagmoma1}.
The evaluation of $\anlyeff$ by two different approaches
was discussed in the text and the results were shown in
Figure~3. 
Further elaborations are made in Ref.~\cite{collar0806v3a1}.
The rest of this reply would focus on $\trigeff$.

\section{II. Discrepancies and Origins}

Ref.~\cite{collar0806v2a1} challenged our
values of $\daqeff \sim 89\%$,
stating that the 
``discriminator threshold of 20~eV''
and
``RMS resolution of pedestal (that is, RT events) of 55~eV''
would imply a DAQ rate ($R$) of $\sim$20k~Hz, based on 
a known relation~\cite{formulaa1}
\begin{equation}
\label{eq::daqrate}
R \sim \frac{1}{4 \tau} ~ 
exp ~ [ - \frac{d^2}{2 ~ \sigma ^2 } ]  ~~,
\end{equation}
where $\tau$ is the shaping time (=6~$\mu$s for the
trigger signal $\sasix$), 
$\sigma$ is the RMS of the pedestal noise
fluctuations 
and
$d$ is the threshold level above the pedestal.
The calculated rate is much higher than the actual DAQ 
rate of $\sim$5~Hz for the ULEGe in the
actual measurement.

This discrepancy originates from the fact that
it is incorrect to use 
these two energy values together in Eq.~\ref{eq::daqrate}.
In the experiment, the ``energy'' of an event is
defined by integrating (the ``Q-mode'') the $\sa12$ signal
within time intervals (-20,51.2)~$\mu$s where t=0 denotes
the trigger instant.
That is, the energy is measured
through the summation of 71.2X20=1424 FADC numbers.
Under this definition, the pedestal (RT) events have
an RMS resolution of 55~eV, after appropriate calibration.
However, the triggering was done by comparing
the $\sasix$ signal  
with a pre-set discriminator level. That is, the
relevant quantity in this operation is the amplitude
of the pulse (the ``A-mode'') $-$ and in fact 
of a different pulse ($\sasix$ instead of $\sa12$, but 
this is a minor point). 
{\it The RMS-resolution of RT events
in Q-mode is not related to the trigger configuration
in A-mode and therefore should not be taken as the $\sigma$ 
of Eq.~\ref{eq::daqrate} which describes the trigger rate.}
It is an experimental optimization
frequently used  (also discussed in Ref.~\cite{formulaa1})
$-$ that the energy definitions are
based on averaging over long periods,
while the triggering schemes make use of 
the instantaneous response of the signal.

\section{III. Evaluation of Trigger Efficiency}

\begin{figure}
%%\vspace*{4ex}
\includegraphics[width=8.5cm]{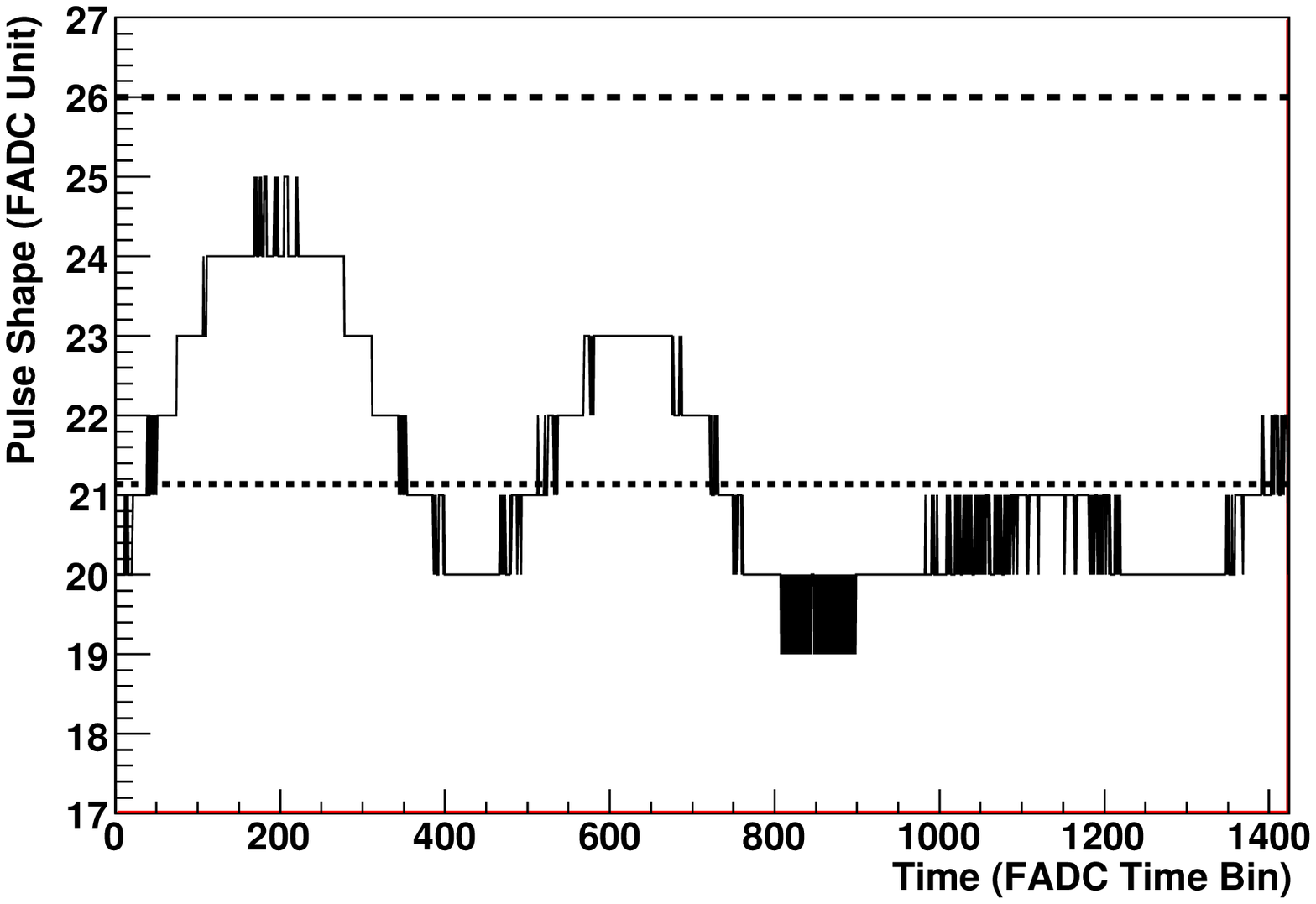}\\
\vspace*{4ex}
\includegraphics[width=8.5cm]{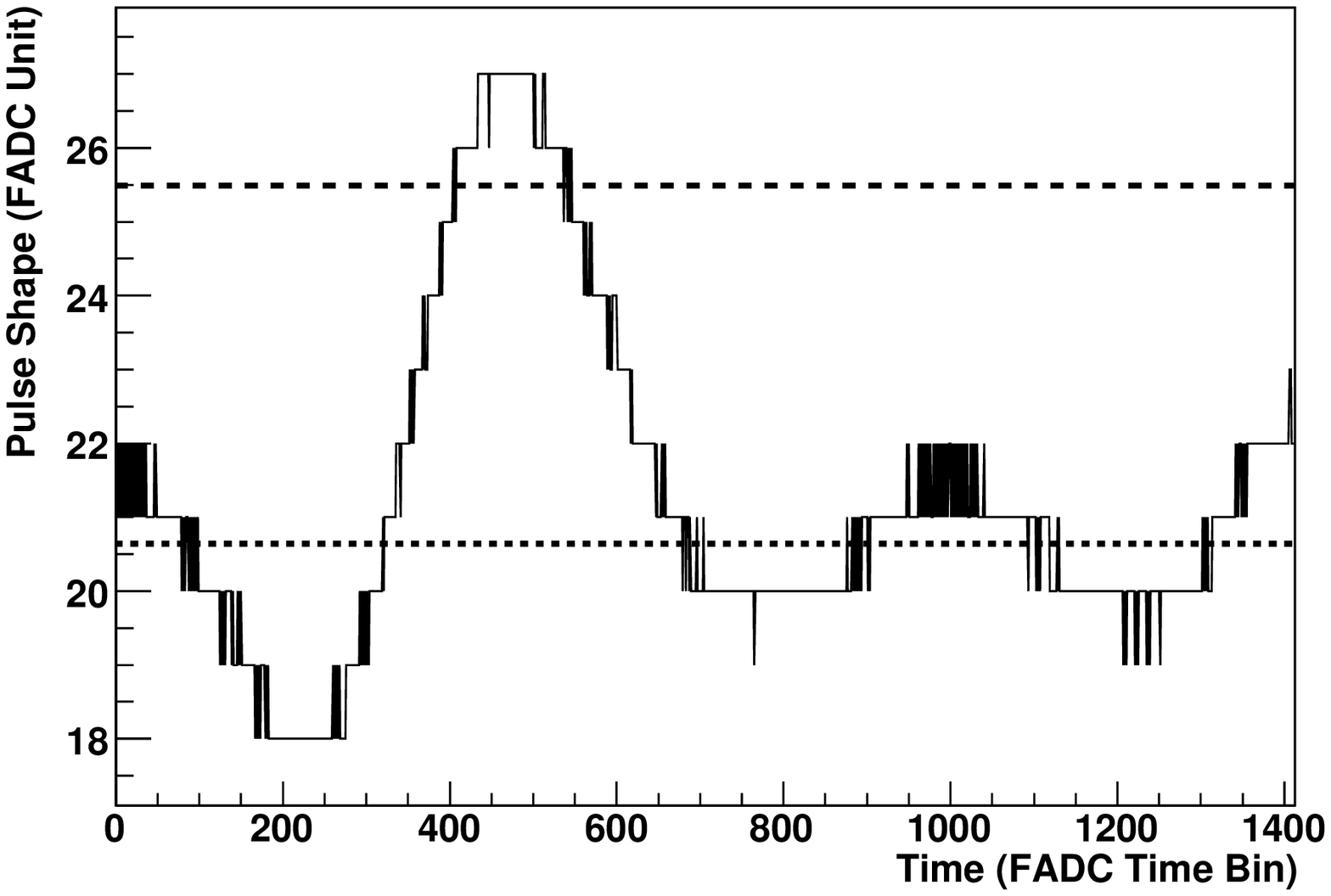}
\caption{
Typical $\sasix$ pulses due to
(a) Top: a random trigger event and 
(b) Bottom: a physics event at 139~eV
in coincidence with anti-Compton signal.
The pedestal mean and the discriminator threshold 
were denoted by dotted and dashed lines, respectively.
}
\label{singleevent}
\end{figure}

Having identified the source of discrepancy which led to
the incorrect conclusions in Ref.~\cite{collar0806v2a1},
we present
the evaluation of $\trigeff$ in what follows.
%%For simplicity and clarity, the data presented are all
%%from a particular channel in the ULEGe.
%%The behaviour of the four channels are similar.
Displayed in Figures~\ref{singleevent}a\&b are typical
$\sasix$ signals for RT and PHY events, respectively.
The amplitude is presented in FADC unit (2~V is equivalent
to 256~FADC~unit).
The PHY-event was measured to be 139~eV in
the Q-mode, and in coincidence with ACV.
Superimposed is the discriminator level
(38$\pm$2~mV in hardware unit,
uncertainties mostly from instabilities over time)
which is 4.8~FADC~unit above the mean value
of the pedestal level.
It can be seen that the RT event is below threshold
while the PHY event is above threshold by
$\sim$ 1.5~FADC unit,
thereby provided a trigger to the DAQ system with good margin.

\begin{figure}
%%\vspace*{2ex}
\includegraphics[width=8.5cm]{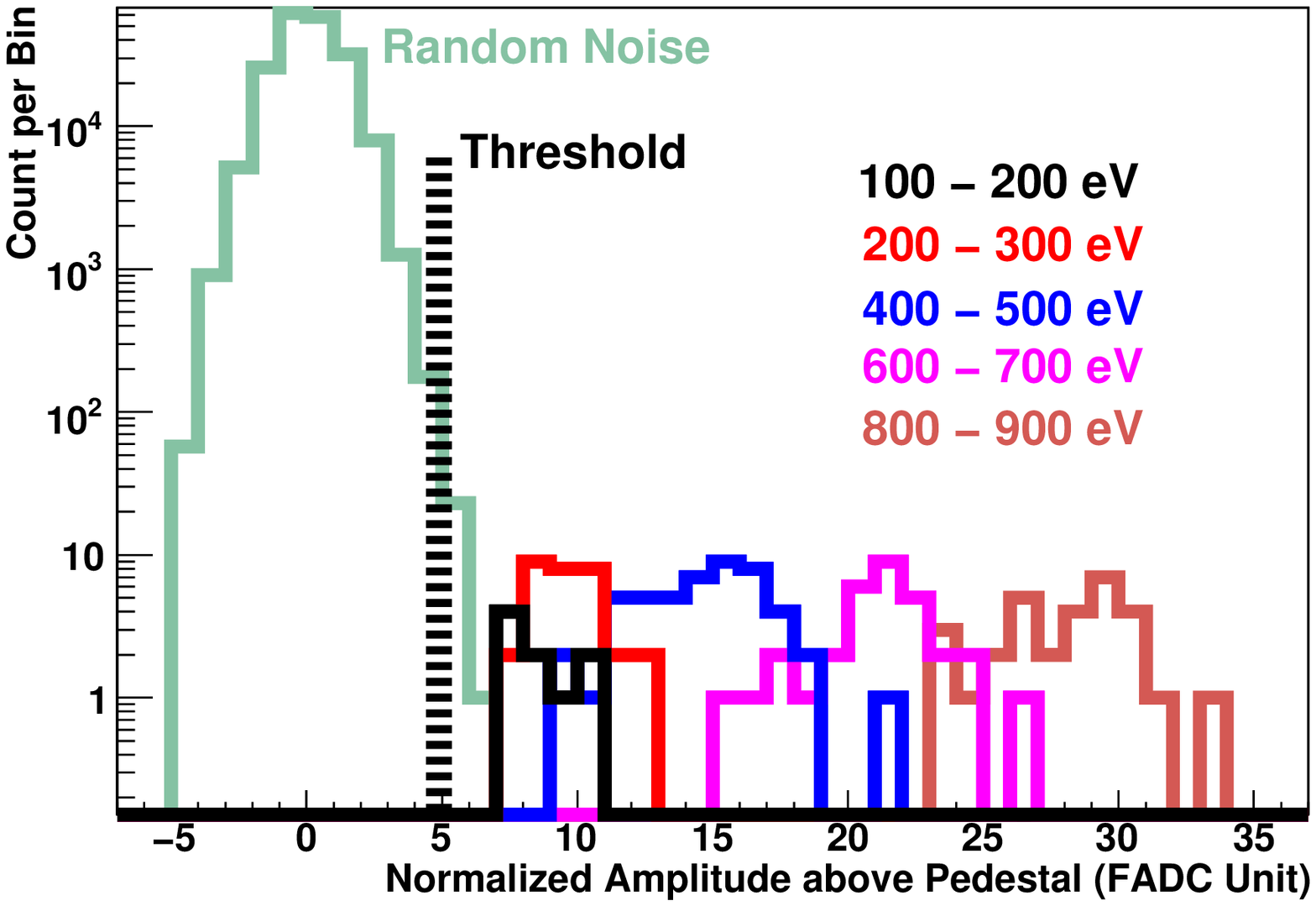}
\caption{
The distributions of noise fluctuation of
RT-events as well as of the maximum amplitudes
of PHY-events in various energy bins.
The discriminator threshold level
is also shown.
}
\label{amode}
\end{figure}

Histograms of large samples of events like those of 
Figures~\ref{singleevent}a\&b is presented in Figure~\ref{amode},
where the horizontal axis is in FADC unit.
The ``Random Noise'' histogram 
corresponds to the distribution of the
amplitudes of every time-bin of the RT-events.
The RMS is 1.1~FADC~unit and represents the
noise fluctuation of the $\sasix$ signal.
The discriminator threshold level is also shown.
Accordingly, one can equate $\tau$=6~${\rm \mu s}$, 
$d$=4.8$\pm$0.3 and
$\sigma$=1.1 in Eq.~\ref{eq::daqrate}, giving
$R \sim 3.1 ^{+6.6}_{-2.2}  ~ {\rm Hz} $.  
%%An upward fluctuation of $\sigma$ by 10\% 
%%would increase $R$ from 0.1~Hz to 0.9~Hz.
The measured rate averaged over the entire DAQ 
periods of about 17 days and for four 
channels together is $\sim$5~Hz.
This is in good agreement.  

\begin{figure} 
%%\vspace*{2ex}
\includegraphics[width=8.5cm]{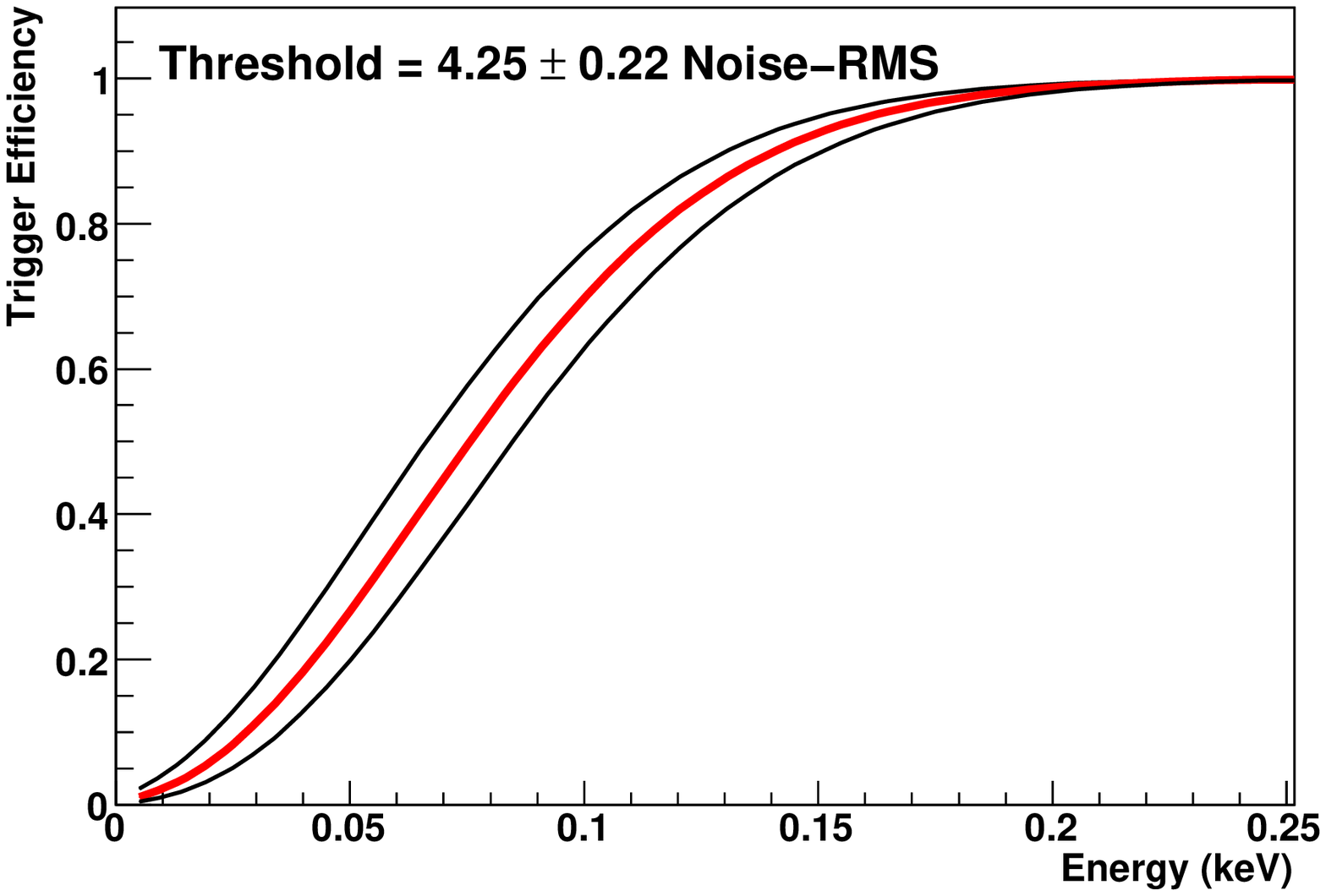}
\caption{
Trigger efficiencies as function of
energy as measured by the Q-mode.
}
\label{trigeff}
\end{figure}

The various
distributions show the
maximum amplitude of PHY-events 
%%selected by ACV-tag~\cite{texono0712}  
at various energy ranges measured by the Q-mode.
All the events above 100~eV
(the relevant range for subsequent analysis) 
exhibit at least 1~FADC~Unit of
margin above threshold.
The trigger efficiencies were then derived using
the maximum amplitude distributions 
of the RT events at E=0 
and the PHY-events between 300~eV to 1000~eV. 
The mean and RMS for the
E=0$-$300~eV regions were evaluated
by interpolation (rather than from actual data)
to avoid biased sampling.
The results are displayed in Figure~\ref{trigeff}.
We note that 
the energy range that provides the most severe constraints to
the dark matter analysis is that of 200-250~eV,
where the trigger efficiency is close to unity.

Using PHY-events as well as data taken
with precision test pulser, it can be 
shown that energy measurements 
with  Q-mode (integration) and 
A-mode (maximum amplitude) are both valid 
and equivalent
for pulses which are 
large compared to the noise fluctuations.
Moreover, the Q-mode measurements are well-behaved and linear
all the way towards the pedestal zero-energy level.
On the contrary, the A-mode 
measurements become inaccurate and 
the calibration is non-linear 
as the energy decreases and approaches zero.
%%Displayed in Figure~\ref{qmode} is the
%%Q-mode energy measurements for the RT events as
%%well as for DAQ events 
%%with maximum amplitude
%%t 0/+1/+2/+3 FADC-units above threshold.
%%The RT-events have an RMS-resolution of 55~eV, 
%%as stated in Ref.~\cite{texono0712}.
%%The events {\it at-threshold} events 
%%correspond to
%%an energy distribution of 89$\pm$42~eV
%%with the  all-event samples. 
%%(There are no PHY-events which are just at-threshold.)
{\it Although the threshold is well-defined in
amplitude (4.8$\pm$0.3~FADC~unit),
the statement ``discriminator threshold at 20~eV''
in Ref.~\cite{texono0712a1} has been misleading
and too simplistic.
It is incorrect to equate
the threshold to a single number in Q-mode
as its energy scale.}
That is,
the trigger efficiency versus energy
curve of Figure~\ref{trigeff}
is not a step-function.
For instance,
$\trigeff = 50\pm30$\% corresponds to
an energy range of about 80$\pm$50~eV.
This is the root of the misunderstanding 
and is corrected in the revised text.

\section{IV. Comparison of KS and Y2L Spectra}

\begin{figure}
\vspace*{5ex}
\includegraphics[width=8.5cm]{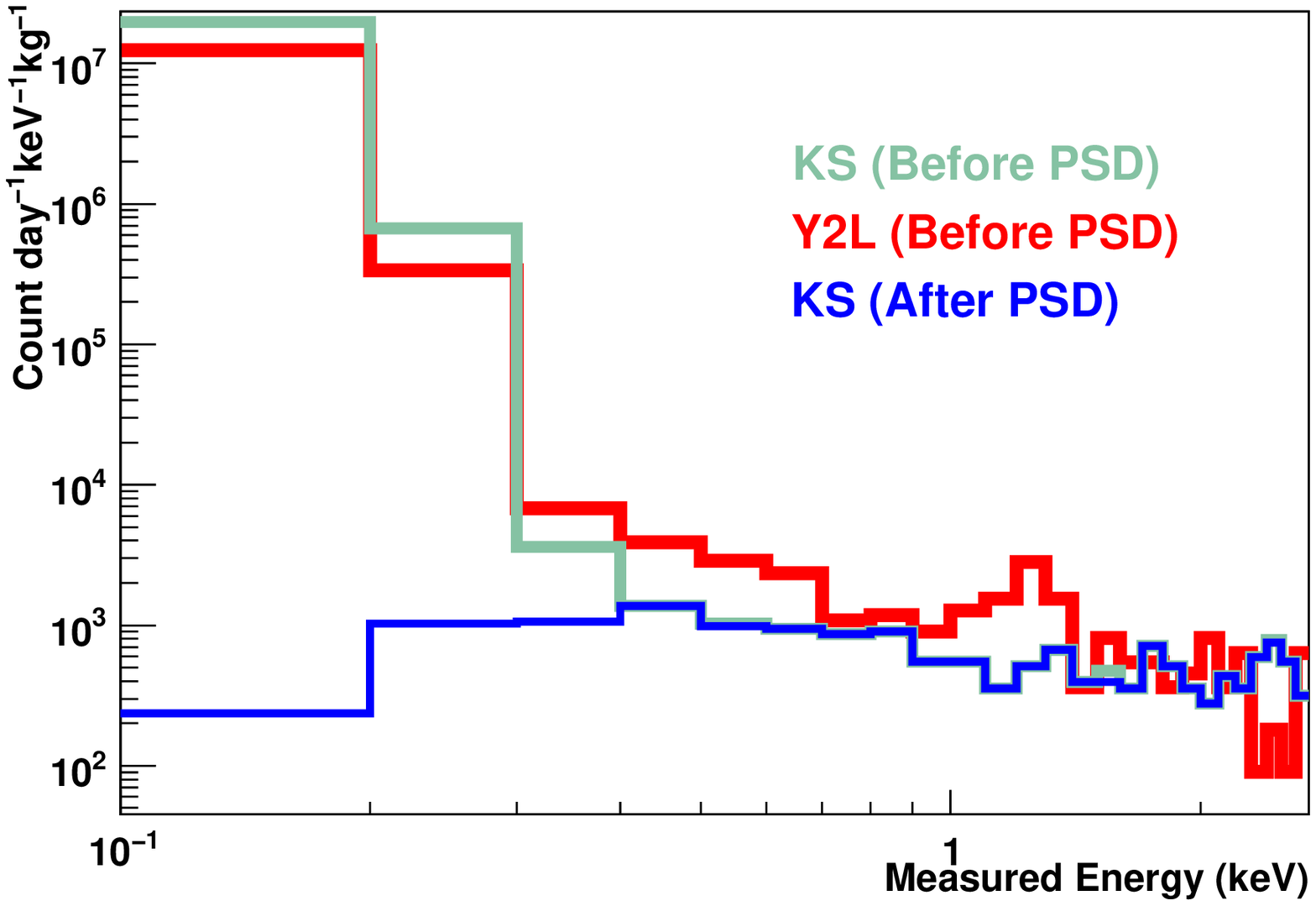}
\caption{
The raw energy spectra
of Y2L and KS, as well as that
after PSD suppression for KS.
}
\label{ksy2l}
\end{figure}

Another point noted in Ref.~\cite{collar0806v2a1} is the comparison
of spectra in Ref.~\cite{texono0712a1} with those taken at the
Y2L underground laboratory in South Korea~\cite{y2la1}.
Such comparisons would not be appropriate. 
Different detectors were 
used even though the specifications
are similar. The electronic modules and noise sources (which
are affected by many ambient conditions) 
are not identical.
The most important difference is that 
we did not make an attempt to use
pulse shape differentiation (PSD) techniques
to suppress ELE-events with Y2L data. 
The spectra shown with Y2L data are always without PSD cuts,
such that the raw ``background'' $-$ dominated by ELE-events $-$
below the noise edge are extremely high.  
DAQ rates in both cases are similar,
at the range 1-10~Hz.
Displayed in Figure~\ref{ksy2l} are the raw spectra
for both KS and Y2L, as well as that after PSD suppression
for KS. It can be seen that the raw spectra are comparable.
The residual differences are due to ambient 
electronic noise and radioactive background
conditions.

We are indebted to 
Profs. F. Avignone, P. Barbeau and J. Collar
for careful reading and critical
comments on Ref.~\cite{texono0712a1}.
Their input
not only expose the parts of the original text 
where the experimental procedures were not 
adequately explained, but
also stimulate us to more thorough 
thoughts on the issue.

%%%%%%%%%%%%%%%%%%%%%%%%%%%%%%%%%%%%%%%%%%%%%%%%%%%%%
% ****** Start of file wimp-reply-v3.tex ****** 

\newpage

\setcounter{figure}{0}

\begin{center}

\large
{\bf 
Appendix II: 
\vspace*{0.2cm}

Selection Cuts and Efficiencies at Threshold $-$\\
Reply to arXiv:0806.1341v3 
}
\normalsize

\vspace*{0.2cm}

H.B.~Li, S.T.~Lin, S.K.~Lin, A.K.~Soma, H.T.~Wong$^*$\\
(TEXONO Collaboration)\\
$^*$ Corresponding Author: htwong@phys.sinica.edu.tw

\end{center}

\vspace*{0.2cm}

Additional questions
were raised
in a recent paper~\cite{collar0806v3a2}
on the experimental concepts and
systematic issues of 
our recent results on
WIMP Dark Matter searches in the 
low-mass ($< 10 ~ {\rm GeV}$) domain with a
Ultra-Low Energy Germanium (ULEGe) detector at 
a threshold of 220~eV~\cite{texono0712a2}.
We provide clarifications and justifications on these
issues, and conclude that 
there are no flaws in our procedures.

\section{I. Comments and Replies}

In a previous version(V2)~\cite{collar0806v2a2},
the authors raised questions on our
DAQ dead time, trigger efficiencies and
apparent inconsistencies between 
the Kuo-Sheng Reactor Laboratory (KS)
and the Yang-Yang Underground Laboratory (Y2L) data.
In an earlier reply~\cite{replyv2a2},
these questions were addressed in details.
No further comments were made along these lines.

A new version(V3) of the comments~\cite{collar0806v3a2}
was subsequently posted, where
questions on the experimental concepts and
various systematic issues were raised.
We responded on these comments in this article.
Section~II deals with the basic concepts and performance
of the PSD cuts and efficiencies, while Section~III
elaborates on the reasons behind the various choices
made in the experiment.
Details of the experimental setup and
data analysis, as well as the definitions
of notations, can be referred to Ref.~\cite{texono0712a2}.

\section{II. Selection Cuts and Efficiencies}

\subsection{A. General Comments}

The objectives of applying
selection criteria (``cuts'') on experimental data
is to reject undesirable ``background'' events and
increase the fraction of ``signal'' events in the
data sample. The events surviving the cuts need not
be all signals, only that usually, the signal-to-background
ratios are enhanced by the selection.

There are much freedom in the choice
of these cuts. There may be effective or
ineffective cuts, but in general, all cuts are valid.
The variables (``figure of merits'' FoM) on which the
cuts are applied are defined to optimize the performance.
The FoMs can be mathematical constructions and
need not correspond to, or be linear to, certain physical variables. 
Once the events are selected
by the cuts, the physical parameters (like energy) can be
derived for these events with different algorithms
which are themselves constructed to give the best resolutions for
these quantities.

When a set of cuts are applied to experimental data, the
corresponding ``selection efficiencies'' must  be evaluated. 
These are the probabilities that potential signal events that will
survive these cuts. The goals of applying cuts are to
suppress background events as much as possible while keeping
the selection efficiencies as large as possible.

\subsection{B. Specific: This Analysis}

In this particular analysis~\cite{texono0712a2},
the ACV and CRV cuts are straightforward.
They are signals from detector 
components other than the ULEGe target, and their efficiencies
were evaluated by random trigger (RT) events.
Their performance are summarized in Table~1 of
Ref.~\cite{texono0712a2}.

It is the PSD-cut which was questioned by Ref.~\cite{collar0806v3a2}.
This cut was applied to the variables
``$\sasix$'' and ``$\sa12$'', as displayed again
in Figure~\ref{fe55}.  Among them, $\sa12$ also provides
an energy measurement, while
$\sasix$ is a mathematical construction to optimize
the performance of signal-vs-background 
differentiation.
In this case, ``background''
corresponds to electronic noise events below
the hardware noise edge.

\begin{figure}
\includegraphics[width=8cm]{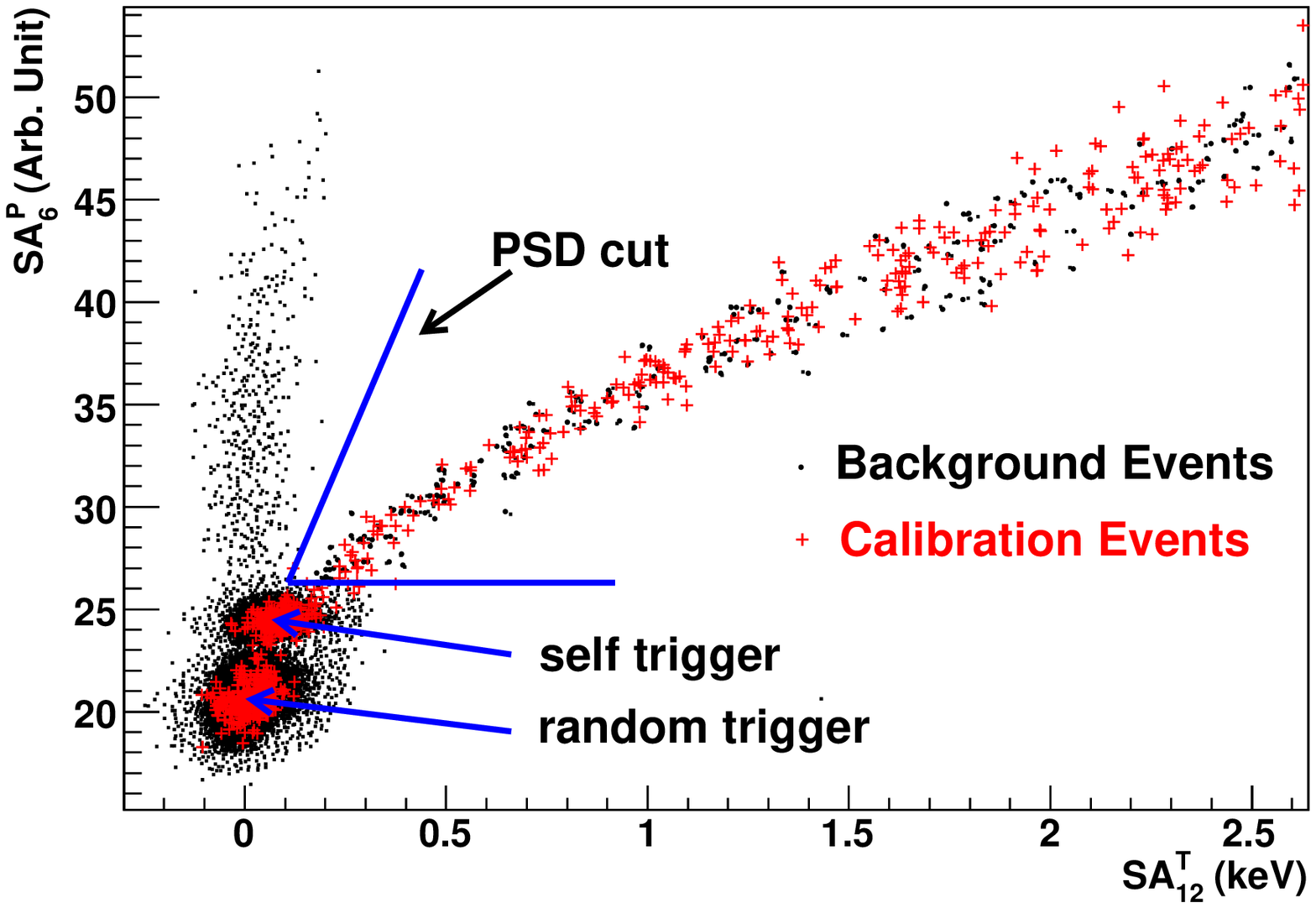}
\caption{
Scattered plots
of the $\rm{SA^{P}_{6}}$ versus
$\rm{SA^T_{12}}$ signals,
for both calibration and physics events
before ACV and CRV cuts.
The PSD selection is shown.
}
\label{fe55}
\end{figure}

{\it Motivations for the choice of $\sasix$:}
The two pulses SA$_6$ and SA$_{12}$ are correlated for
physics signals. The correlations are different and 
less strong for noise-events.
The conceptual idea behind the PSD cut is that,
given the energy of an event is known through
measurements by $\sa12$, 
positive fluctuations of the 
SA$_6$-pulses in physics-induced events
at a particular time interval 
and amplitude ranges can be expected.
This is probed by the PSD cut.
The selected time interval 
is optimized for energies near threshold 
($\rm{< 300 ~ eV}$), such that this
interval no longer corresponds to the amplitude-peaks 
of SA$_6$ for events at higher energy.
Therefore, $\sasix$ being non-linear with energy does
not jeopardize the validity of the cut. The reasons behind 
the non-linear behaviour is well-understood $-$ and in fact
intentional.

{\it Is the PSD cut arbitrary?}. NO. 
The application of the PSD cut  is based on genuine 
understanding on the behaviour of
the detector hardware. Such PSD techniques $-$ correlations of
two signals at two shaping times and, alternatively,
of the full and partial integrations of the signals $-$ 
are well-established
ones at higher energy in the case of $\alpha$/$\gamma$ separation
in many detector systems.

\begin{figure}
\includegraphics[width=8cm]{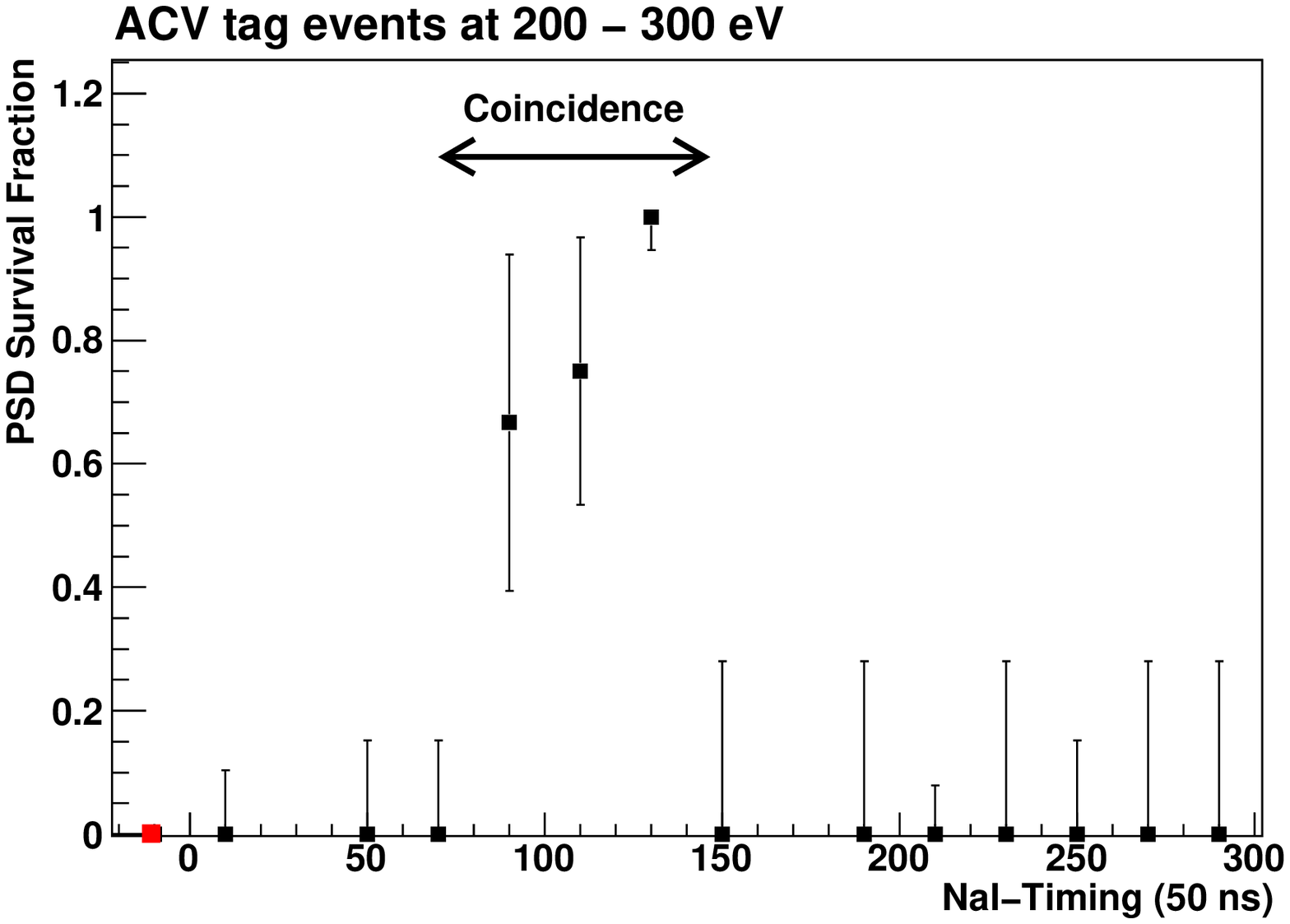}
\caption{
The survival fraction of events at E=200$-$300~eV 
with an ACV (Anti-Compton) tag
versus the relative timing between the ACV signals and the
ULEGe triggers. Overlaid are the actual coincidence time 
interval between the ACV and the ULEGe systems 
derived independently from hardware timing. 
The data point at negative time is due to 
events without ACV-tags, and corresponds to
$( 1.7 \pm 0.3 ) \times 10^{-4}$.
}
\label{acvtag}
\end{figure}

Displayed in Figure~\ref{acvtag} are the
survival fraction ($f$) of 
events at E=200$-$300~eV with an ACV (Anti-Compton) tag
versus the relative timing between the ACV signals and the
ULEGe triggers. Overlaid are the actual coincidence time interval
between the ACV and the ULEGe systems 
determined independently from hardware timing.
The coincidence window is about 3~$\mu$s.
This is defined by  
(a) the 6~$\mu$s shaping time output SA$_6$ from ULEGe 
which provided the trigger timing; and 
(b) the increased time-jitters at low energy.
The ACV and ULEGe signals outside this range 
are accidentals and uncorrelated.

It can be seen that
{\it ONLY} ULEGe events in correct
coincidence with the ACV-tags give a substantial value of $f$.
The survival fraction for events without ACV-tags
(denoted by the data
point at negative time in Figure~\ref{acvtag}) $-$
predominantly due to electronic noise at this energy $-$ 
is $( 1.7 \pm 0.3 ) \times 10^{-4}$.  
This proves that the PSD cut is successfully devised
and is performing its intended functions
of ``suppressing electronic
noise events and selecting the physics-induced events''.

%%The triggered events 
%%below 300~eV 
%%are mostly noise events and 
%%the PSD cut suppressed them
%%by about  $\sim 10^{-4}$. 
{\it Selection Efficiencies:}
The PSD-efficiency for physics-induced events
($\psdeff$) is related to 
the measured survival fraction $f$ by 
\begin{equation*}
f ~ = ~  
\frac{( \psdeff  *  P + f_N * N )}{P+N} ~ , 
\end{equation*}
where
$P$ and $N$ are, respectively,
the numbers of physics-induced and noise events
in correct coincidence with ACV, while $f_N$ is
the survival fraction for events triggered by
electronic noise. 
In this experiment, 
$f$ and $(P+N)$ are measured
quantities, $f_N \sim 10^{-4}$ at 200$-$300~eV
and even less at lower energy, 
while the {\it average} $N$ can be
evaluated from the non-coincidence samples
where $P=0$.
It can be seen that, in general,
$ \psdeff  >   f $ so long as there
are finite fraction of noise events in 
the ACV-tagged sample.

Statistics are limited in the present analysis 
since only {\it in situ} data were used.
The number of noise events in the 
coincidence ACV-tagged sample
is finite but
has large uncertainties.
Accordingly, the more appropriate approach 
is to take zero noise-background ($N=0$),
giving rise to the
assignment of $ \psdeff  =   f $.
The subsequent upper limits derived
would therefore be conservative 
and less constraining ones.
The variations of $\psdeff$ with energy 
under this assignment 
are displayed in Figure~3 of 
Ref.~\cite{texono0712a2}.

\section{III.  Experiment's Choices}

Inevitably,
an experiment has to choose 
various tools from the available pool
in the course of its analysis. We explain 
why such choices were made for
the cases raised in Ref.~\cite{collar0806v3a2},
and illustrate
the sensitivities to the physics
results if alternative schemes 
would had been selected instead. 

\subsection{A. $^{55}$Fe spectrum for Selection Efficiency}

It was explicitly stated in Ref.~\cite{texono0712a2}
that the flat $^{55}$Fe spectrum 
at low energy is an assumption.
Under this assumption, the PSD selection efficiencies
were derived. The results agree well with those obtained
by the more rigorous approach with ACV-tagged events.
The good agreement suggests validity of the assumption
at the present level of accuracy.
Accordingly, it is justified to add these
results to further constrain (to reduce uncertainties of) 
our knowledge of $\psdeff$. This is
the approach adopted in Ref.~\cite{texono0712a2}.

If the data from the $^{55}$Fe spectrum 
would be ignored altogether and
the PSD efficiencies would be derived exclusively
from the ACV-tagged events, 
$\psdeff$ at 
E=200$-$300~eV would move
from (0.65$\pm$0.06) to (0.61$\pm$0.08).
The 50\% efficiency line would correspond to
224~eV instead of 216~eV, while
the $\csnospin$ limits 
(in units of $10^{-39} ~ {\rm cm^2}$ throughout in this Section)
at $\mwimp = 5 ~{\rm GeV}$ 
would increase (become less constraining)
from 0.81 to 0.88.

\subsection{B. Quenching Factor}

\begin{figure}
\includegraphics[width=8cm]{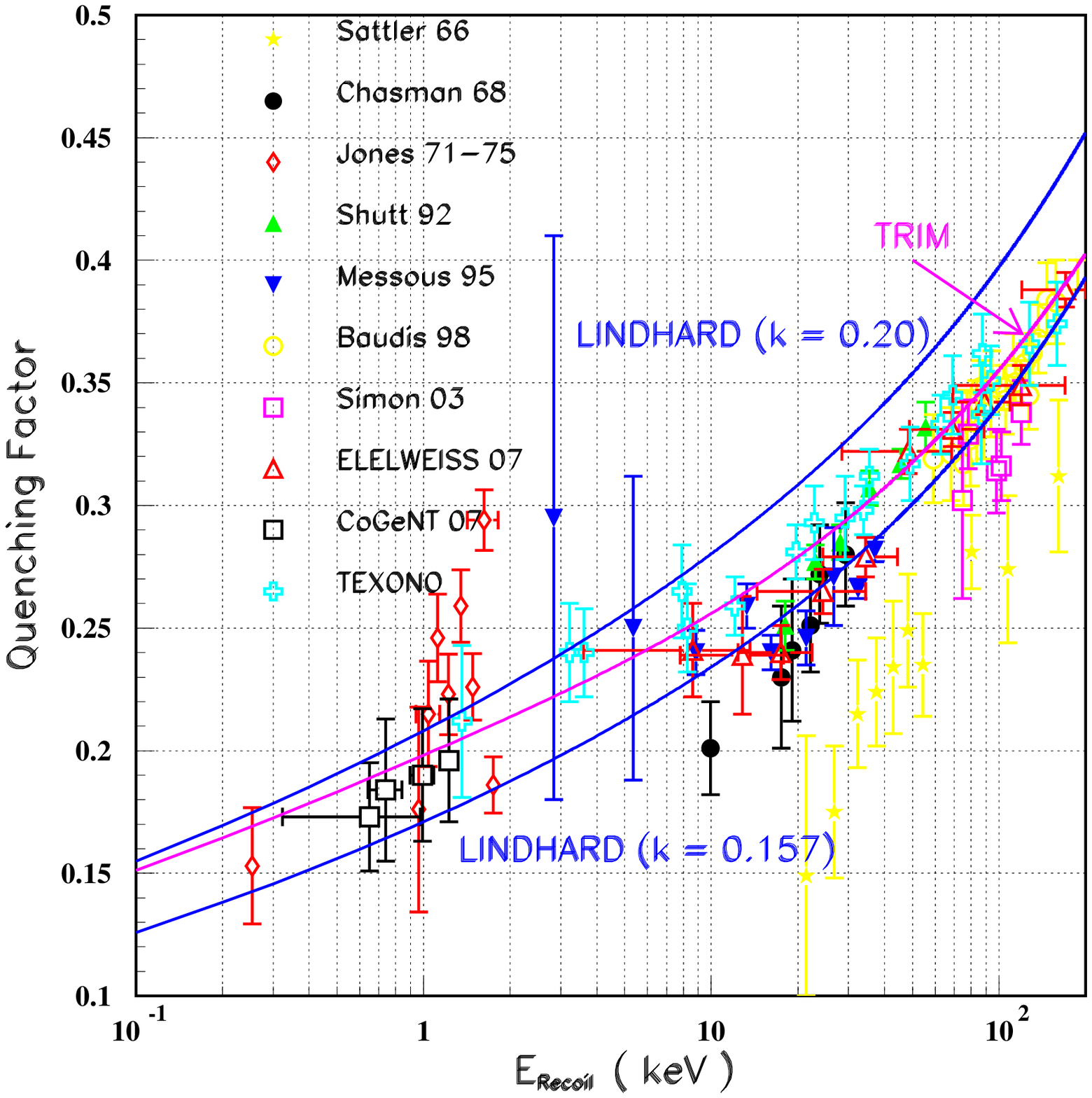}
\caption{
A compilation of all quenching factor (QF) measurements 
on germanium, with
calculations from the TRIM software~\cite{trima2}
as well as by the Lindhard model~\cite{lindharda2}
under two parametrizations (k=0.20 and 0.15)
overlaid.
}
\label{qfdata}
\end{figure}

A compilation of all quenching factor (QF) measurements 
on germanium
is given in Figure~\ref{qfdata}. Overlaid are
calculations from the TRIM software~\cite{trima2}
as well as by the Lindhard model~\cite{lindharda2}
under two parametrizations (k=0.20 and 0.157).
Both schemes have been adopted in various
CDM experiments.
It can be seen that the TRIM results explain
well the QF measurements at both low and high 
energy. Accordingly, we chose to use this scheme
in our analysis. The QF values are less
than those evaluated with the Lindhard (k=0.20) model,
and hence would give rise to 
more conservative results.

If Lindhard (k=0.20) would be used, the
QF at 1~keV recoil energy will be increased
from 0.20 to 0.21. 
The QF uncertainty estimations of 0.006
in Ref.~\cite{texono0712a2}
can account for this deviation.
This alternative choice will only have 
minor effects on the exclusion limits, 
decreasing it (becoming more constraining) 
from 0.81 to 0.80
at $\mwimp = 5 ~ {\rm GeV}$ 

\subsection{C. Constructing Exclusion Plots}

The unbinned ``optimal interval method'' as
formulated in Ref.~\cite{yellina2} was adopted 
to derive the exclusion limits. 
The unbinned formalism allows the use of
all available information in the background 
spectra  and was used in other CDM experiments 
like CDMS and XENON. 
NO background profile was assumed or subtracted, 
which is also a conservative approach.
The sensitivities at low $\mwimp$ 
under this scheme are driven by
the absence of counts between 198~eV and 241~eV.

An alternative method
would be to place the background
events in different energy bins and follow the formalism 
of Ref.~\cite{binneda2}. %%(??).
For instance, 
choosing 50-eV bins for E$>$100~eV (thereby
deliberately filling the hole at 200$-$250~eV),
the $\csnospin$ limit at $\mwimp = 5 ~ {\rm GeV}$ 
would increase (become less constraining)  %%by ??\% 
from 0.81 to 1.20. This reduction in
sensitivities is expected 
since data binning involves loss of information.

We conclude that our choices in these three
aspects of the experiment are justified.
The sensitivities of the
physics results (exclusion upper limits)
are dominated by the statistical
uncertainties of the background spectra.
The potential effects on them
are minor if alternative schemes
would have be chosen instead.

\end{document}